\documentstyle[prb,eqsecnum,aps]{revtex}

\begin{document}

\title{ \bf
Exact, Complete, and Universal Continuous-Time Worldline Monte Carlo
Approach to the Statistics of Discrete Quantum Systems }

\author{
N. V. Prokof'ev, B. V. Svistunov, and I. S. Tupitsyn}

\address{
Kurchatov Institute, 123182 Moscow, Russia }

\maketitle

\begin{abstract}
We show how the worldline quantum Monte Carlo procedure, which
usually relies on an artificial time discretization, can be
formulated directly in continuous time, rendering the scheme exact. 
For an arbitrary system with discrete Hilbert space, none of the 
configuration update procedures contain small parameters. 
We find that the most effective update strategy involves the motion 
of worldline discontinuities (both in space and time), i.e.,
the evaluation of the Green's function. Being based on local updates only, 
our method nevertheless allows one
to work with the grand canonical ensemble and non-zero winding 
numbers, and to calculate any dynamic correlation function as easily
as expectation values of, e.g., total energy. The principles found 
for the update in continuous time generalize to any continuous 
variables in the space of discrete
virtual transitions, and in principle also make it possible to
simulate continuous systems exactly.

\end{abstract}

\bigskip

               \section{Introduction}
\label{sec:1}

Quantum Monte Carlo (MC) simulation is the most powerful
available method, if not the only one, of obtaining accurate results for
complex systems, where analytic solutions are not
possible and exact diagonalization methods do not work because of the
enormous Hilbert space. However, most MC schemes are far from ideal,
and suffer from significant shortcomings. These include (see, e.g.,
the most recent review article Ref.~\cite{Sandvik-review})

a) systematic errors due to artificial time discretization,
      which in most schemes scales as $(\Delta \tau )^2$, where 
      $\Delta \tau $ is the time slice width; 
      
b)  restriction of the simulation to the zero winding number sector
      $M=0$ (a configuration in which world lines connect 
      the initial state  $\vert \alpha_1 , \alpha_2 , \dots , 
      \alpha_L \rangle $ at $\tau =0$ to the final state
      $\vert \gamma_1 , \gamma_2 , \dots , 
      \gamma_L \rangle $ at $\tau =\beta $, with the set 
      $\{ \gamma_i \}$ being obtained by
      cyclically  permuting $\{ \alpha_i \}$ $M$ times (and all
      topologically equivalent configurations), is said 
      to have a winding number $M$). Such a restriction results in 
      systematic errors too, which however
      vanish with increasing system size. Also, one loses the ability
      to study topological excitations in the system, e.g., vortices or
      supercurrent states; 
      
c) working with a fixed number of particles
      $N= \mbox{const}$ (canonical ensemble); 
      
d) critical slowing down problem, which arises close to a
      second-order phase transition. This problem is closely related
      to constraints (b) and (c), and is indicative of inefficient procedures
      used to update configurations with large length scales;

e) slow accumulation of statistics when calculating
      correlation functions of operators not present in the 
      initial Hamiltonian, e.g., the Green's function; 
      
f) small acceptance rates in update procedures. These may be due to
      from small parameters present in the formulation of the MC scheme,
      or systems described
      by Hamiltonians with different energy scales (e.g., when 
      the hopping matrix element $t$ is much smaller than the  
      typical potential energy change $U \gg t$), or the necessity of global
      Metropolis updates, which arise in certain cluster-update algorithms.
      
g) anomalous dependence of the computation time on system size 
      (due to self-averaging effects in the thermodynamic limit,
      the computation time required to achieve given accuracy is expected 
      to be system-size independent); 
      
h) notorious sign problem, which emerges when the configuration weight
      is not positive definite. Since we do not see any reasonable
      solution of the sign problem in the general case, in what follows we 
      exclude it from the discussion.

To eliminate some of these shortcomings, a number of different MC 
schemes were developed. Unfortunately, none of the 
existing schemes succeeded in solving all of them 
(leaving the sign problem aside) in the general case:
there are extremely efficient algorithms which are far from 
universal, while the efficiency of existing universal algorithms is far 
from high for a large number of problems.

The standard worldline 
algorithm is based on imaginary time discretization and
utilizes the small parameter $t \Delta \tau \ll 1$ in an approximate
treatment of noncommuting operators in the Hamiltonian, 
known as Trotter break-up \cite{Suzuki,Hirsch81}. Physical intuitiveness
and easy programming probably make this method the one most 
widely used. On another hand, its weak points
range over the whole list from (a) to (f), the most severe ones being 
(e) and (f).  

In the worldline algorithm, one describes the
configuration by specifying the system state $\vert \alpha_k \rangle $
at all time slices
$\tau_k=k \: \Delta \tau$, where $k=0,1,\dots ,K_\beta$ and 
$t_{K_\beta}=1/T \equiv \beta$. The system state is then conventionally 
defined
in the basis set in which the potential energy of the system is diagonal,
i.e., in the site representation. Consider, as a typical example, 
the Hamiltonian of interacting particles on a lattice 
\begin{equation}
H= -t\sum_{<ij>} a_i^{\dag}a_j + \sum_{ij}U_{ij}n_in_j \;,  
\label{Ham}
\end{equation}
where $a_i^{\dag}$ creates a particle on site $i$, $t$ is the hopping matrix
element, $n_i=a_i^{\dag}a_i$, and $<ij>$ denotes nearest-neighbor sites. 
From now on, we call points in time 
at which the system changes state ``kinks." 
The typical separation in time between two
adjacent kinks on the same site is of order $1/t$ and independent 
of $\Delta \tau$, so that
for small $\Delta \tau$ there are some $1/(t\Delta \tau ) \gg 1$ time
intervals between them. The acceptance rate of the variation
suggesting  creation of a new kink - antikink
pair is proportional to the square of a small parameter,
$(t\Delta \tau )^2$. On the other hand, when
the MC procedure suggests shifting an
already existing kink to the nearest point in time, the corresponding
variation of the configuration is accepted with probability $\sim
O(1)$. Thus, on average, by selecting different random time slices, 
it takes some $1/(t\Delta \tau )^2$ attempts to
create a new kink - antikink pair and $1/(t\Delta \tau)$ attempts to move
a kink to the nearest position in time. Still, the updated 
configuration is only slightly different from the previous one, and 
expectation values calculated before and after the variation are strongly
correlated. An uncorrelated contribution of the given configuration
fragment is obtained by shifting the kink a distance of order $1/t$,
which requires some $1/(t \Delta \tau )^3$ operations, since the kink
shift process is diffusive in nature. This means that the 
autocorrelation time in the standard worldline algorithm grows as
$ \propto (\Delta \tau)^{-3}$ even in the absence of critical slowing down.
Since all update procedures are local, the algorithm is subject to
critical slowing down near the transition temperature.

In order to calculate the Green's function ${\cal G}(i, \tau )$, 
two worldline
discontinuities are inserted at time slices $ \tau_1=0$ and 
$ \tau_2=\tau $ (in other words, one extra worldline is
inserted in or removed from the interval $[\tau_1 , \tau_2 ]$ )
\cite{Hirsch81,Kashurnikov}. One then
probes different configurations using standard update procedures and
collects statistics in a $d$-dimensional histogram, which describes
the spatial separation $i$ between the discontinuities. The
length of the time interval is then changed, and the same calculation
is repeated. 
One MC step, i.e., the number of update operations performed between 
successive ``measurements," whereupon another point is included in the
statistics of the calculated quantity, is proportional to $L^d \beta $,
where $L^d$ is the number of lattice sites considered.
Thus it requires about $L^d \beta $ operations to include 
only one point in the $(d+1)$-dimensional spacetime histogram
for $G$. 

In a way, the standard worldline procedure of 
calculating ${\cal G}(i,\tau )$ has an anomalous dependence on 
$N$ and $\beta $,
since it takes at least $(N \beta )^2$ operations to update the whole
histogram (typically $G$ decays in space and time,
and large scale behavior requires much more computation). 
We note, that winding numbers and the grand canonical ensemble average
can be incorporated, in principle, in the worldline algorithm. It is
sufficient to consider separate contributions to 
the statistics of ${\cal G}(i,\tau ) $ when
 $i=ML$ and $\tau =n\beta $ with integer $M$ and $n$.
However, in practice, only small systems at rather high temperatures
can be considered using this algorithm.
 
The determinant method based on the Hubbard - Stratonovich transformation
\cite{Blankenbecler,Hirsch85,White89} also uses the discrete-time 
Trotter break-up, and thus becomes more and more inefficient
due to long autocorrelation times when $\Delta \tau \to 0$
(points (a) and (f) above). It has an important advantage over
the worldline method in calculating the Green's function,
since it works with the grand canonical ensemble. However with
increasing system size, the calculation time scales as $L^3$ 
(point (g)) and some of the procedures become ill-conditioned at 
low temperatures.

Another technique allowing Green's function calculations is called 
Green-function MC (or, more generally, the projection-operator 
method) \cite{Ceperley}. It is applicable at zero temperature
only, and the final result for ${\cal G}(i,\tau =0)$ depends on 
the trial wave function (we are not aware of whether it is possible 
to calculate the time dependence of ${\cal G}(i,\tau )$ by this method).

The stochastic series expansion (SSE) technique 
\cite{Sandvik-review,Sandvik91,Sandvik92}, 
which stems from the Handscomb's method \cite{Handscomb}, 
relies on the direct Taylor expansion of the statistical operator.
This scheme is {\it exact} (contains no systematic errors).
SSE has clearly demonstrated that time discretization is an artificial 
trick that is not at all necessary for MC simulation. Since an 
elementary update in the SSE scheme is equivalent to roughly 
$1/(t\Delta \tau )^3$ updates in the standard worldline method, 
it results in a significant drop in computation time for 
high-precision calculations. The rest of the problems, i.e., (b)-(f),
survive in the SSE approach (point (f) still applies, because by expanding
in powers of the full Hamiltonian one has to compare weights corresponding
to the kinetic- and potential-energy terms, and if, e.g.,  $U \gg t$ in the 
Hamiltonian (\ref{Ham}), then small acceptance rates appear in the update
procedures). Still, away from the transition point, for large systems at 
low temperature, for which $U \sim t$, the SSE method is 
superior in evaluating basic thermodynamic properties like the total 
energy and density - density or current - current correlation functions.

A qualitatively new class of extremely efficient MC schemes 
\cite{Swendsen} has been developed in recent years
\cite{Evertz,Wiese,Kawashima94,KG,Kawashima96,Beard}. These schemes are
based on the so-called loop cluster update (LCU) algorithm, which performs
nonlocal updates for worldline loops with sizes as large 
as the system itself. Apart from solving the problem of critical
slowing down, it also allows one to work in the grand canonical ensemble
and with nonzero winding numbers. From this method we learn that problems
(b), (c) and (d) can be circumvented. Unfortunately, the LCU algorithm,
as far as we know, is not universal. It applies to spin systems and to
hard-core Hubbard models, but was never formulated for the general 
lattice Hamiltonian, like interacting soft-core bosons with 
arbitrary $U_{ij}$, arbitrary density (chemical potential), or on-site
disorder. Another shortcoming of the LCU, which in a sense can also
be called nonuniversality, is that it does not admit of a universal code
\cite{com1}. It should be noted that LCU allows for considerable 
generalization to the cases of external magnetic field and disorder,
but generally speaking the cost is lost efficiency because of 
the exponentially small acceptance rates for large loops \cite{com2}.

It is shown in Ref.~\cite{Farhi} how to build a path integral in continuous 
time for quantum systems in a discrete basis. The configuration is 
specified by transition times and system states before and after the 
transition. Within this description one can formally think 
about taking the limit $\Delta \tau \to 0$ in the standard approach. 
However, to implement this description one has to formulate the update
process. In the standard worldline algorithm, one of the basic
procedures is generation of new kink-antikink pairs when
system evolution on a given site changes from $\vert \alpha \rangle \to
\vert \alpha \rangle \to \vert \alpha \rangle $ to
$\vert \alpha \rangle \to
\vert \gamma \ne \alpha \rangle \to \vert \alpha \rangle $.
In the continuous limit, the acceptance rate for such a variation 
vanishes as $(\Delta \tau )^2$, and thus the problem of qualitatively
new update principles arises.

Recently, two independent continuous time schemes utilizing the ideas
of Ref.~\cite{Farhi} were developed \cite{Beard,our}. 
Beard and Wiese \cite{Beard} find that with the 
LCU algorithm, one can go directly to the continuous-time limit 
$\Delta \tau \to 0$, thus rendering the LCU algorithm {\it exact}. 

The general solution to the problem of configuration update 
in continuous time is found in Ref.~\cite{our}. 
The resulting continuous-time worldline (CTWL) method is exact, and
like SSE, is $1/(t\Delta \tau )^3 $ times more efficient than 
finite-$\Delta \tau$ local schemes. It also completely eliminates
problem (f), since none of the procedures relies on small parameters
(potential energy is accounted for in the exponent, and one does not
have to weight relative contributions to the statistics of the 
potential and kinetic energy terms, as happens in SSE).

In its original formulation, the CTWL approach did not solve problems
(b) - (e), and was tested only on a simple single-particle Hamiltonian
\cite{our}. In the present article, we present a complete description of the
CTWL approach to the statistics of arbitrary many-particle system with
discrete Hilbert space. We demonstrate that it enables one to solve
problem (e) in a physically intuitive way by formulating the {\it local}
update procedures in terms of the motion of two worldline discontinuities
(in what follows we call them ``worms") in space and time, i.e., in terms of
a calculation of the Green's function. During one MC step
(consisting of $N \beta /t$ operations) the whole histogram
for ${\cal G}(i,\tau )$ is updated, which means that $G$ is calculated
as efficiently as, say, the total energy, and is not affected by
point (g). Since ${\cal G}(i=[M_xL_x,\: M_yL_y, \dots ], n \beta )$ with
integer $M_x, \: M_y, \dots $ and $n$ describes a system with
 $n$ extra particles and winding numbers $ \{ M \}$, we are working
in the grand canonical ensemble. This solves problems (b) and (c).

Closer examination of the loop building rules \cite{Wiese,Beard}
for the Heisenberg Hamiltonian shows their remarkable similarity
to the evolution of an extra worldline segment.
The crucial difference is that only closed loops are considered 
by the LCU algorithm, while our scheme considers
all the intermediate configurations as well, and utilizes them 
for the Green's function calculation. Working in the 
extended configuration space, which includes discontinuous worldlines,
we use local Metropolis-type \cite{Metropolis} updates only. However, 
when discontinuities annihilate, and we return to the configuration space
of closed worldlines, the net result of the update is of global character.
Since CTWL with ``worm" updates effectively mimics single-loop LCU, we may
hope that it possesses all the remarkable features inherent in LCU, and in
particular that it solves, or at least softens, the problem (d).

To summarize, we propose a method which is {\it exact}, 
{\it complete} (allows calculation of any
correlation function), and {\it universal} (applies to arbitrary
quantum systems with discrete Hilbert space, and enables one to write a 
unified code, that is simultaneously
applicable to lattice bosons and arbitrary spins, with 
arbitrarily long-range interactions and disorder). The sign problem 
now becomes the only stumbling block on the way of making quantum MC 
an ``ideal" computational tool for studying complex systems.

This paper is organized as follows. In Section \ref{sec:2} we formulate  
the general principles of the continuous-time worldline approach.
In Section \ref{sec:3} we introduce the update procedures
that we find to be the most effective and sufficient
for simulation of the quantum statistics of many-particle systems. 
In Section \ref{sec:4} we demonstrate the advantages of the new method by
presenting some of results that cannot be obtained by any other
MC approach: the Green's function and the critical index
of the 1D boson Hubbard model at the quantum critical point,
and the low-energy properties of the strongly-disordered
Bose glass phase.
In Section \ref{sec:5} (and Appendix A) we discuss the feasibility
of increasing the efficiency of our method in the case of long-range 
interactions, and consider the feasibility of generalizing of the method
to continuous systems.

               \section{General principles}
\label{sec:2}

Let $H_0$ and $V$ be the diagonal and off-diagonal parts of the
Hamiltonian $H$ in a chosen representation corresponding to the full set
$\{ \alpha \}$ of eigenstates of $H_0$, with 
$H_0 \mid \alpha \, \rangle = E_{\alpha} \mid \alpha \, \rangle $.
The statistical operator can then ordinarily be related to the Matsubara
evolution operator $\sigma$ in the interaction picture, i.e., we
write $e^{- \beta H} = e^{- \beta H_0} \sigma $, with 
\begin{equation}
\sigma  =  1 - \int_0^{\beta} d \tau \; V(\tau) + \ldots  
 +  (-1)^m \int_0^{\beta} d \tau_m  \cdots
\int_0^{\tau_2} d \tau_1 \: V(\tau_m) \cdots V(\tau_1) +
\ldots \; ,  \label{rel1}
\end{equation}
where $V(\tau) = e^{ \tau H_0} V e^{- \tau H_0}$. Without loss of generality
and in accordance with typical forms of Hamiltonians of interest, $V$ can be
written as a sum of elementary terms $Q_s$, whose action on any function
from the set $\{ \alpha \}$ results in another function from this set: 
\begin{equation}
V = \sum_s Q_s \; , \;\;\;\;
Q_s \mid \alpha \, \rangle = - q_{\gamma \alpha}(s) \mid \gamma \, \rangle
\;\;\;\;\;\;(\gamma = \gamma (s,\alpha )) \;.  \label{Q}
\end{equation}
Since $V$ is Hermitian, for any $s$ in the sum (\ref{Q}) there exists 
an $s^{\prime}$ such that $Q_{s^{\prime}} = Q^{\dag}_s$. We rewrite
Eq.\ (\ref{rel1}) in components (below $E_{\alpha \gamma} \equiv E_{\alpha}
- E_{\gamma}$): 
\begin{eqnarray}
\sigma_{\alpha \gamma} & = & \delta_{\alpha \gamma} + \sum_s \int_0^{\beta}
d \tau \: q_{\alpha \gamma}(s) e^{\tau E_{\alpha \gamma}} + \ldots  \nonumber
\\
& + & \sum_{s_1, \ldots , s_m} \int_0^{\beta} d \tau_m \cdots
\int_0^{\tau_2} d \tau_1 \: q_{\alpha \nu}(s_m) e^{\tau_m E_{\alpha \nu}}
\cdots q_{\lambda \gamma}(s_1) e^{\tau_1 E_{\lambda \gamma}} + \ldots \; .
\label{rel2}
\end{eqnarray}
Note that there is no additional summation over the indices of the
intermediate complete sets (labeled by Greek letters), since these are
defined in a unique way by configurations of $(s_1, s_2,\ldots , s_m)$.

We confine ourselves to the case of finite-range interaction, which is
defined by the requirement that for each term $s_1$ of elementary operators 
$\{ Q_s \}$ there exists only a finite number of terms $s_2$ for which the
condition 
\begin{equation}
\left[ \, Q_{s_1}(\tau_1), \, Q_{s_2}(\tau_2) \, \right] = 0  \label{comm}
\end{equation}
is not met. In the case of finite-range interaction, the structure of the
series (\ref{rel2}) is drastically simplified, the simplification being of
crucial importance for practical realization of our algorithm. From 
(\ref{comm}) it follows that up to an irrelevant change in the indexing 
of energies and matrix elements, one can ignore the chronological order of 
$Q_{s_1}(\tau_1 )$ and $Q_{s_2} (\tau_2 )$ in the evolution operator. This
suggests representing a general
term of the series (\ref{rel2}) in the following form. First, we introduce
the notion of a ``kink of type $s$," which is characterized by a time 
$\tau$, a matrix element $q_{\alpha \gamma}(s)$, 
and a diagonal-energy difference $E_{\alpha \gamma}$. 
The former two we will refer to as parameters of the
kink. It is essential that (i) to obtain parameters of a kink one need not
know explicitly the whole state $\mid \alpha \, \rangle$, or $\mid \gamma \,
\rangle$ -- local information is enough; (ii) to specify a particular 
structure
of a term in Eq.\ (\ref{rel2}), including the chronological order of all
noncommuting operators, it is enough to specify for each kink its associated
neighbors, i.e., the noncommuting kinks nearest in time.

Now our goal is to describe in general terms a stochastic process that 
directly evaluates Eq.\ (\ref{rel2}). For simplicity, we assume that all 
$q_{\alpha \beta}(s)$ are positive real numbers. (In many particular 
alternative cases, a straightforward generalization is possible, but usually 
at the expense of convergence.) Summations and integrations in 
Eq.\ (\ref{rel2}) then can be regarded, up to a normalizing factor, 
as an averaging over the statistics of different configurations of kinks, 
each configuration being defined by a certain number of kinks of certain 
types, their associations and particular positions in imaginary time. 
The Monte Carlo process should examine these statistics by generating 
different kink configurations in accordance with their weights. The global 
process will consist of a number of elementary subprocesses, each being 
responsible for certain modifications of a particular type. 

An update procedure of a general type should involve 
subprocesses of creation and annihilation of kinks. Clearly,
the qualitative difference between discrete- and continuous-time
QMC schemes is associated with processes of just this kind.
To introduce the general principles of construction of subprocesses
that change the total number of kinks, we consider some particular
(but still rather rich) class of elementary transformations (which seems 
to be sufficient for all practical purposes). By an elementary transformation
we mean a subprocess which either only creates or only
annihilates a certain number of kinks. The set of elementary
subprocesses can be decomposed into self-balanced creation - annihilation 
pairs. Our task then is to specify the structure of creation and
annihilation subprocesses, and to derive the balance equation that
would guarantee that the statistics generated by each pair of 
subprocesses does really correspond to that introduced by 
Eq.\ (\ref{rel2}). 

Let some subprocess create $n$ kinks of given type $s_1, s_2, \ldots ,
s_n$, the temporal positions of the kinks being specified by the
$n$-dimensional vector $\vec \tau  = \{ \tau_1, \tau_2, \ldots, \tau_n \}$.
In the most general case, the creation procedure involves two steps.

First, one suggests to create $n$ new kinks at 
$\vec \tau  \in \Gamma$, where $\Gamma$ is a certain region in the
$n$-dimensional space of times $\tau_1, \tau_2, \ldots, \tau_n $.
The probability density $W(\vec \tau)$ of choosing a 
given $\vec \tau$ is, generally speaking, arbitrary, provided
$W(\vec \tau)$ is nonzero at every physically meaningful configuration
of kinks.

In the second step, one either accepts (with probability
$P_{\mbox{\scriptsize acc}}(\vec \tau)$) or rejects the suggested 
modification.

The annihilation procedure is much simpler. The $n$ kinks of given
type $s_1, s_2, \ldots , s_n$ and with $\vec \tau  \in \Gamma$ are
either removed (with probability $P_{\mbox{\scriptsize rem}}(\vec \tau)$)
or remain untouched.

The equation of balance for the given pair of subprocesses reads
\begin{equation}
A_0 \, p_{\mbox{\scriptsize c}} \, W(\vec \tau) \, 
P_{\mbox{\scriptsize acc}}(\vec \tau) \, d \vec \tau \; - \; 
dA_n(\vec \tau ) \, p_{\mbox{\scriptsize a}} \, 
P_{\mbox{\scriptsize rem}}(\vec \tau)  \; = \; 0 \; \; .
\label{balance}
\end{equation}
Here $A_0$  ($A_n(\vec \tau)$) is the probability (probability density)
of finding a configuration without the specified
$n$ kinks (with the specified $n$ kinks at the given $\vec \tau$).
We have also introduced the probabilities $p_{\mbox{\scriptsize c}}$
and $p_{\mbox{\scriptsize a}}$ of addressing the creation
and annihilation subprocedures. In the next section we will see
how it can turn out quite naturally that these probabilities do not
coincide.

The statistical interpretation of Eq.\ (\ref{rel2}) implies
\begin{equation}
\frac{dA_n(\vec \tau)}{A_0} \; = \; d \vec \tau \, \prod_{j=1}^{n}
q(s_j) \, e ^{ \Delta E_j  \tau_j } \; ,
\label{stat}
\end{equation}
where $q(s_j) \equiv q_{\alpha_j \beta_j}(s_j)$ and 
$\Delta E_j \equiv E_{\alpha_j } - E_{\beta_j }$.
Combining (\ref{balance}) and (\ref{stat}) we obtain the necessary
and sufficient condition for the pair of subprocesses to be
self-balanced:
\begin{equation}
\frac{W(\vec \tau) \, P_{\mbox{\scriptsize acc}}(\vec \tau) }
{P_{\mbox{\scriptsize rem}}(\vec \tau)} \, = \, R(\vec \tau) \, , \; \; \; 
R(\vec \tau) \, = \, 
\frac{p_{\mbox{\scriptsize a}}}{p_{\mbox{\scriptsize c}}} \, \prod_{j=1}^{n}
q(s_j) e ^{ \Delta E_j \tau_j } \; .
\label{balcond}
\end{equation}
Given $W(\vec \tau)$, the condition (\ref{balcond}) is satisfied, e.g., 
by the following obvious choice of $P_{\mbox{\scriptsize acc}}$ and 
$P_{\mbox{\scriptsize rem}}$:
\begin{equation}
P_{\mbox{\scriptsize acc}}(\vec \tau)  \: = \: \left\{ 
\begin{array}{ll}
R(\vec \tau) / W(\vec \tau) \; ,
\mbox{~~~~~~~~~~ if $R(\vec \tau) <  W(\vec \tau) $} \;\; , \\
\;\;\;\;\;\;1 \; ,  \mbox{~~~~~~~~~~~~~~~~~~~ otherwise} \;\;\;\;\; ,
\end{array} \right.
\label{P_acc}
\end{equation}
\begin{equation}
P_{\mbox{\scriptsize rem}}(\vec \tau)  \: = \: \left\{ 
\begin{array}{ll}
W(\vec \tau) / R(\vec \tau) \; ,
\mbox{~~~~~~~~~~ if $R(\vec \tau) > W(\vec \tau) $} \;\; , \\
\;\;\;\;\;\;1 \; ,  \mbox{~~~~~~~~~~~~~~~~~~~ otherwise} \;\;\;\;\; .
\end{array} \right.
\label{P_rem}
\end{equation}
From (\ref{P_acc}) it can be seen that there is a certain reason for 
choosing $W(\vec \tau) \propto R(\vec \tau)$, as in this case
$P_{\mbox{\scriptsize acc}}$ becomes independent of 
$\vec \tau$, and the accept - reject
decision can be made before suggesting a particular configuration,
thus saving computational time. However, if the structure of the
function $R(\vec \tau)$ is complicated, the numerical generation
of the corresponding distribution will be very expensive. In this case
it is better to take $W(\vec \tau) \propto \tilde{R}(\vec \tau)$, 
where $\tilde{R}(\vec \tau)$ is some ``coarse-grained''
approximation to $R(\vec \tau)$ with a simple form.

We do not consider here a general theory of subprocesses
that do not change the number of kinks, since it is basically
the well-known theory of taking multidimensional 
integrals by standard Monte Carlo procedures. Particular examples of such 
subprocesses can be found in Ref.~\cite{our} and in the next section.

The foregoing approach does not involve any explicit truncation of
the series (\ref{rel2}). One might wonder, however, what the effect of
implicit truncation in the practical realization of the process would be,
due to the finite size of the computer memory. To this end 
we note that even for simulations of many-particle systems, where the
typical number of kinks $N_{\scriptsize kink}$ (that is, the typical
number of terms in the series (\ref{rel2})) that contribute to the final
result is really large, and one might expect the memory/accuracy problem,
the effect can be easily made absolutely negligible. Indeed, from 
the Central Limit Theorem, it follows that the number of kinks in
significant configurations has a Gaussian distribution with the peak
at $\bar{N}_{\scriptsize kink}$ and a half-width of order
$\sqrt{\bar{N}_{\scriptsize kink}}$ (cf. Ref.\ \cite{Sandvik91}).
If one just reserves at least twice as much memory
as necessary to describe the configuration
with $\bar{N}_{\scriptsize kink}$ elements, then during a computation 
spanning the age of the Universe, the system will not fluctuate 
to states which cannot be fit into memory. The implicit truncation 
error thus can be made astronomically small.

            \section{Update procedures}
\label{sec:3}

\subsection{Kink motion}
\label{kink-motion}

Let us first consider update procedures that are straightforward
generalizations of those known in the discrete-time worldline 
algorithm, and work with closed trajectories only. The simplest
process involves transformations that do not change the number
of kinks, but change their types, time positions, and temporal 
ordering, \cite{our}
\begin{equation}
\langle  \alpha  \vert \: Q_{a_1}(\tau_{a_1})
Q_{a_2}(\tau_{a_2}) \dots Q_{a_n}(\tau_{a_n})
\: \vert  \gamma   \rangle  \longrightarrow
\langle  \alpha  \vert \: Q_{b_1}(\tau_{b_1})
Q_{b_2}(\tau_{b_2}) \dots Q_{b_n}(\tau_{b_n})
\: \vert  \gamma  \rangle \;.
\label{move-1}
\end{equation}
The number of operators involved in the transformation, their
types, and time positions are not constrained, except that the two
configurations have nonzero weight. Obviously, one could suggest
many different realizations of Eq.~(\ref{move-1}), and some
might work more efficiently than others, depending on the
system. Here we describe the procedure called ``kink motion";
other procedures have too much in common to be described separately,
and allow trivial modifications.

To move a kink, we first select it at random from the list of existing
kinks and decide on the time interval to be considered. Suppose
that we have chosen a transition described by $(Q_o, \tau_o )$. We
then find kinks of the same type that are nearest in time (both to 
the left and to the right of $\tau_o$), i.e., $Q_o$ or $Q_o^{\dag }$,
and consider their times $\tau_1 <\tau_o$ and $\tau_2 > \tau_o$ as
the boundaries of the time ``window" transformed by this procedure
(in certain configurations at high temperature, it may happen that
$(\tau_1 ,\tau_2 )= (0, \beta )$ ). It is allowed to have any number 
of kinks of different types $Q_a \ne Q_o, Q_o^{\dag }$ within 
$(\tau_1 , \tau_2 )$. Thus the typical initial configuration has 
the form
\begin{equation}
\dots \bigg\vert_{\tau_1} \; Q_{a_1}(\tau_{a_1})
Q_{a_2}(\tau_{a_2}) \dots Q_o(\tau_o) \dots Q_{a_n}(\tau_{a_n})
\; \bigg\vert_{\tau_2} \dots \;,
\label{move-2}
\end{equation}
(as explained above, one has to consider only those kinks which do not
commute with $Q_o$).

The second step is to analyze all possible configurations obtained
from (\ref{move-2}) by removing $Q_o$ from point $\tau_o$ and inserting 
it at arbitrary $\tau ' \in (\tau_1 , \tau_2 )$. We keep the time
positions and the chronological ordering of all the other
operators $Q_{a_1},Q_{a_1}, \dots, Q_{a_n}$ untouched. 
The new position of the selected kink $Q_o$ in time is decided
according to the statistical weight  of the final configuration
as defined by Eq.~(\ref{rel2}). This is done in complete analogy
with the classical MC procedure of taking multidimensional integrals.

The acceptance rate of the kink motion procedure is unity, since
the differential measure of the initial configuration is zero.
In this way, all noncommuting kinks in the Hamiltonian (except 
kink - antikink pairs, which are dealt with in the next subsection)
can change places. In dimensions $d>1$, the kink motion procedure
must be supplemented with a ``local loop" procedure, which generates
small loops in real space, e.g., by replacing
$Q_{i\to i+g_1}(\tau_1 )Q_{i+g_1\to i+g_1+g_2}(\tau_2 ) \longrightarrow 
Q_{i\to i+g_2}(\tau_3 )Q_{i+g_2 \to i+g_1+g_2}(\tau_4 ) $,
where $g_1,g_2$ are the nearest neighbor indices.

\subsection{Creation and annihilation of kink - antikink pairs}
\label{kink-anncre}

In this subsection we make use of the general theory  of
Sec.~\ref{sec:2} and explain how the elementary procedure of creation and
annihilation of kink - antikink pairs is organized in practice.
An important new principle realized in our algorithm is the possibility
of selecting different update procedures with certain probabilities
(see also Appendix A).
These probabilities, $p_a$ and $p_c$, are at our disposal, and if necessary, 
can be used to ``fine tune" the efficiency of the MC process as a whole.
The most natural starting point for the update 
is to address at random some configuration fragment. It can be characterized
by the kink $Q_o(\tau_o)$, or by the system state $\vert 
\alpha (i_o) \rangle$ between the two adjacent kinks that change 
this state (in computer memory, all $\vert \alpha (i_o) \rangle $ 
between kinks are assigned labels; the configuration itself
is described as a linked graph by specifying nearest-neighbor
associations (in space and time) between the labels). We choose the 
latter variant and address site labels. Thus the probability of 
applying an update procedure to a given fragment is $\propto 
1/N_{\mbox{\scriptsize lab}}$, where $N_{\mbox{\scriptsize lab}}$ is
the total number of labels characterizing the initial configuration. 
By inserting (deleting) $n$ extra kinks, we increase (decrease) 
$N_{\mbox{\scriptsize lab}}$ by $ \sum_{j=1}^{n}m_{Q_j}$ where
$m_{Q_j}$ gives the number of states changed by the kink $Q_j$.
Thus the ratio $p_a/p_c$ in Eq.~(\ref{balcond}) is proportional to
$N_{\mbox{\scriptsize lab}}/
(N_{\mbox{\scriptsize lab}}+\sum_{j=1}^{n}m_{Q_j})$
when addressing the creation of $n$ kinks, and 
$(N_{\mbox{\scriptsize lab}}-\sum_{j=1}^{n}m_{Q_j})/
N_{\mbox{\scriptsize lab}}$
when addressing the annihilation procedure.

To fix the values of $p_a$ and $p_c$, we count the number 
of possible kink - antikink processes that can be applied to a given
fragment. This number is denoted by $N_{\mbox{\scriptsize proc}}$.
The simplest choice is then to assign equal weight, 
$1/N_{\mbox{\scriptsize proc}}$, to all possibilities.
For example, if we consider a model with the nearest-neighbor
hopping in 1D, then there are three possibilities for the site
state $\vert \alpha (i) \rangle$: to insert 
$Q_{i \to i+1}Q_{i+1 \to i}$ or $Q_{i \to i-1}Q_{i-1 \to i}$
and to delete a pair of kinks that change this state to the left 
and to the right in time, provided they form a kink - antikink
pair (i.e., are of the $Q_{i\pm 1 \to i}Q_{i \to i \pm 1}$ type).
In this case $m_{Q_j}=2$ as well, and we finally have
\begin{equation}
{p_a \over p_c} = {N_{\mbox{\scriptsize lab}} \over
N_{\mbox{\scriptsize lab}}+4 } ~~~~\mbox{(creation)}\;; ~~~~~~~~
{p_a \over p_c} = {N_{\mbox{\scriptsize lab}}-4 \over
N_{\mbox{\scriptsize lab}}  } ~~~~\mbox{(annihilation)} \;.
\label{pa-pc}
\end{equation}
Obviously, in the thermodynamic limit and at low temperature, these
ratios are very close to unity. Again, this is only a particular
example; other choices may prove to be more efficient under
certain conditions.

Once the configuration fragment and update procedure are selected,
we proceed along the lines described in Sec.~\ref{sec:2}. Here we
would like to comment on the choice of probability density
$W(\vec{\tau })$. It would be perfect from the acceptance rate 
point of view to take $W(\vec{\tau }) \propto R(\vec{\tau })$.
However, this can turn out to be a very expensive procedure. To 
illustrate the point, consider a configuration fragment of length
$\tau_{l,r}=\tau_r-\tau_l$. 
Due to the large interaction radius between particles,
an effective field acting on updated states can change many times
during $\tau_{l,r}$. If the number of time slices thus induced
on the interval $(\tau_l,\tau_r) $ is $N_{\tau_{l,r}} \gg 1$, then
complete parametrization of the $R(\vec{\tau })$ function will
require calculation of the $N_{\tau_{l,r}}(N_{\tau_{l,r}}+1)/2$
partial probabilities, according to the number of ways one can
distribute two kinks among $N_{\tau_{l,r}}$ time subintervals.

The solution of the problem lies in choosing $W(\vec{\tau }) =
W (\overline{E},\vec{\tau })$, where
$W(\overline{E},\vec{\tau })$ is
an analytic function with the same properties as  
$W(\vec{\tau })$, controlled by a parameter $\overline{E}$ that
is used to minimize the variance of $\vert 
W (\overline{E},\vec{\tau }) - R(\vec{\tau })
\vert $. The most obvious physical choice of $\overline{E}$ is the 
mean field potential acting on the updated states during $\tau_{l,r}$ from
the rest of the system
\begin{equation}
e^{-\overline{E} \tau_{l,r}} = R(\tau_l,\tau_r) \;,
\label{meanfield-1}
\end{equation}
\begin{equation}
W (\overline{E},\vec{\tau })=
{e^{-\overline{E} (\tau_2-\tau_1)} \over I } \; , ~~~~~~~
I=\int_{\tau_l}^{\tau_r}  d\tau_2
\int_{\tau_l}^{\tau_2}  d\tau_1
e^{-\overline{E} (\tau_2-\tau_1)} \; .
\label{meanfield-2}
\end{equation}
One immediately recognizes in 
$W (\overline{E},\vec{\tau })$ the
statistics of the kink - antikink pair in the biased two-level system
\cite{our}, which, through the mean-field definition of the bias 
energy $\overline{E}$, most closely approximates 
the local statistics of kink - antikink pairs in a real system.

The procedures described in the last two subsections represent
a direct generalization of local procedures already known in the 
discrete-time worldline method. Their continuous-time
versions are, however, only specific realizations of a much 
wider class of possible procedures, thus making the overall
CTWL scheme more flexible. 

\subsection{Creation - annihilation, jump, and reconnection 
procedures for worldline discontinuities}
\label{worms-jr}

Up to now, we have considered procedures for working with closed worldlines.
These are sufficient to simulate quantum statistics in the canonical
ensemble and in the $M=0$ sector. To overcome this essential
drawback, and to calculate the Green's function, one usually
introduces an extra worldline segment and simulates 
quantum statistics in the presence of two worldline discontinuities
at points $(i_1,\tau_1)$ and $(i_2,\tau_2)$. This process is
highly inefficient, because one has to probe all degrees of freedom
in the configuration (numbering roughly $ \sim L^d \beta$)
to collect statistics for only two
extra degrees of freedom. In practice, this method was never used 
to calculate Green's function  in large systems, e.g., with
$L^d \beta \sim 10^{4}$. The solution we find for this problem is
in considering the two worldline discontinuities to be real dynamic
variables in the Hamiltonian, which are allowed to move
through the configuration both in space and time. It turns out that
this motion can be arranged to be ergodic, and probes all possible
system states.  One can even completely ignore all the other
update procedures, such as moving other kinks and working with 
kink - antikink pairs, probably at the expense of being less efficient,
but still remaining accurate, complete, and universal. Below we describe
the details of update procedures with worldline discontinuities
(``worms"), which were first introduced in Ref.~\cite{our}.

We start with the general expression for the Matsubara Green's function
(see, e.g., Ref.~\cite{AGD}) in the interaction picture
\begin{equation}
{\cal G}(i,j,\tau_1,\tau_2 ) =  -e^{\beta \Omega} 
Tr\:\bigg[ e^{-\beta H_0} T_{\tau }
\big(  a_i(\tau_1 ) a^{\dag }_j (\tau_2)  \sigma \big) \bigg]  \;,
\label{Green}
\end{equation}
where $T_{\tau }$ is the $\tau$-ordering operator, which was explicitly
written before in defining the Matsubara evolution operator $\sigma$ 
in Eq.~(\ref{rel1}); $\Omega $ is the grand canonical potential. 
To be specific, we assume here that $H_o$ is diagonal in the site
representation; in the general case, one might imagine that the index
$i$ refers to some parametrization of eigenstates of $H_o$.
Since we now work in the grand canonical ensemble, the Hamiltonian contains
an extra term
\begin{equation}
-\mu N \equiv -\mu \sum_{i}n_i \;,
\label{chempot}
\end{equation}
where $\mu$ is the chemical potential. Formally, the only difference
between the statistics given by Eq.~({\ref{rel2}) and
the Green's function (\ref{Green}) is that we have two 
extra kinks, $a_i (\tau_1)$ and $a^{\dag }_j (\tau_2)$. 
Hence one has the possibility of calculating the Green's function
in a {\it unified process}, together with standard thermodynamic
averages (``energy," for the sake of brevity). To this end, it is necessary 
just to work in an extended configuration space, where two classes of 
configurations are present: (i) with continuous worldlines, and (ii) with 
two worldline discontinuities, corresponding to the kinks 
$a_i (\tau_1)$ and $a^{\dag }_j (\tau_2)$. (Clearly, configurations
of class (i) contribute to ``energy," while those of the class (ii)
contribute to the Green's function.) The transitions between the two
classes are performed by the processes of creation and annihilation of
the kinks $a_i (\tau_1)$ and $a^{\dag }_j (\tau_2)$, in accordance
with the general balance principles Eq.~(\ref{balcond}). For
computational purposes, it is reasonable to redefine
the Green's function by a trivial scaling transformation
$a_i \rightarrow \eta^* a_i \, , \, 
a^{\dag }_j \rightarrow \eta a^{\dag }_j$, where the constant
$\eta$ is adjusted to produce the optimal acceptance (rejection) 
probability.

Alternatively, one can arrive at the above scheme by the standard 
trick of introducing a source to the configuration action $S$
(the notation $\eta$ for the source is chosen deliberately):
\begin{equation}
\int_0^\beta d\tau V(\tau )  \longrightarrow 
\int_0^\beta d\tau V(\tau ) + \sum_{i}\int_{0}^{\beta } d\tau
(~\eta^{*}_i(\tau )~ a_i(\tau )+\eta_i(\tau )~a^{\dag}_i(\tau ) ) \;,
\label{source}
\end{equation}
and defining the Green's function as a functional derivative of the 
generating functional (the partition function with the source)
\begin{equation}
{\cal G}(i,j,\tau_1,\tau_2 )= -{1 \over Z}\: {\delta^2 Z \over \delta
\eta_i(\tau_1 )\delta \eta^{*}_j(\tau_2 ) } 
\bigg\vert_{\eta ,\eta^{*} \to 0} \;.
\label{Green2}
\end{equation}
The numerical procedure equivalent to the variational derivative 
in the limit $\eta \to 0$ means that only configurations with (i) zero 
and (ii) two worldline discontinuities are included in the statistics.
Confining ourselves to just these configurations, we do not have to
deal any longer with infinitesimally small $\eta$ , and can choose $\eta$
to be a certain finite constant. (This is crucial for any realistic 
computational process, since $\eta \to 0$ clearly means that the 
time of accumulation of statistics goes 
to infinity.) Indeed, a particular value of $\eta$ just 
defines the relative weights of classes (i) and (ii), thus changing 
the relative norm of the Green's function with respect to ``energy"  
by the known factor of $\mid \eta \mid^2$. (Incidentally, one may pay no 
attention at all to the normalizing statistics for the Green's function, 
as the norm can ultimately be fixed by the condition 
${\cal G}(i,i,\tau, \tau +0 )= - \mbox{density}$.)

A typical configuration with two ``worms" is shown in Fig.\ 1
(``live" picture taken from the computer). To update
it we apply the following transformations:
\begin{itemize}
\item
{\sl Creation and annihilation of two worldline discontinuities}\\
We delete a pair, $a_i(\tau_1) a^{\dag}_i(\tau_2) $ or 
$a^{\dag}_i(\tau_1) a_i(\tau_2)$, when discontinuities happen to 
meet at the same site $i$ and there are no other kinks between them
that can change the state of $i$. The only difference between this 
and the kink - antikink procedure is that 
now we transform only a single-site state, thus $m_Q=1$. The annihilation 
procedure addresses the pair of worms, but the creation procedure
(which makes sense only when there are no worms) addresses the 
randomly selected configuration fragment label. The ratio of probabilities
$p_a/p_c$ to address update procedures that transform the same
configuration fragment back and forth is now
\begin{equation}
{p_a \over p_c} = N_{\mbox{\scriptsize lab}} ~~~~\mbox{(creation)}\;;
~~~~~~~~
{p_a \over p_c} = N_{\mbox{\scriptsize lab}}-2 ~~~~\mbox{(annihilation)} \;.
\label{wormspa-pa}
\end{equation}
This ratio is macroscopically large, which is obviously unpleasant for 
the computational process. However, we have 
$R(\tau ) \sim \mid \eta \mid ^2 $, with the freedom of choosing $\eta$.
By setting $\mid \eta \mid ^2 \sim 1/ 
\langle N_{\mbox{\scriptsize lab}} \rangle $, where
$\langle N_{\mbox{\scriptsize lab}} \rangle $ is the average number of 
labels in the configuration, we obtain an update procedure that is not 
based on small parameters (in practice, any rough estimate like $(L^d
\beta )$ for $\langle N_{\mbox{\scriptsize lab}} \rangle $ is sufficient).
The rest is done in exactly the same manner as described in 
Sec.~\ref{kink-anncre}.
\item
{\sl Jump }\\
This update procedure is illustrated in Fig.\ 2. We select one of the 
worldline discontinuities and suggest shifting it in space by
inserting an  ordinary kink (hopping operator) 
to the left (in time) of the annihilation operator and to the right of the
creation operator. As a result, the worm ``jumps" to another site.
The number of kinks changes by one in this procedure, but $p_a/p_c$
is unity, because we address it upon the availability of worms, and
not according to the number of labels. Also, since we are dealing with
only one extra kink here, the structure of the $R(\tau )$ function
(see Sec.~\ref{sec:2}) is much simpler, and we choose 
$W(\tau) = R(\tau )/\int d\tau R(\tau )$. The integral is over the
time interval of the updated fragment. The opposite procedure is called
an ``anti - jump."
\item
{\sl Reconnection}\\
Formally, this update procedure, which is shown in Fig.\ 3,
is technically identical to the ``jump," but now an extra kink
is inserted to the right of the annihilation operator and to
the left of the creation operator. We still
distinguish between them, because in the jump procedure the corresponding
particle trajectories do not exchange places, while they do
so in the reconnection update. Figure 3 makes it clear that
we have effectively reconnected worldline segments of different 
trajectories. 
Note that in fermionic systems, any reconnection/antireconnection
procedure results in a change of the configuration sign.
\item
{\sl Shift in time}\\
The motion of worldline discontinuities in time is essentially
the kink motion process (see Sec.~\ref{kink-motion}).
Suppose that we have decided to shift an annihilation kink $Q_o=a_i$.
The only difference from the scheme (\ref{move-2}) 
is in the definition of the updated time interval. 
Its boundaries $(\tau_1 ,\tau_2)$ now
correspond to the time positions of the nearest left and right
neighbors (kinks) of any type that operate on the same state $i$.
Of course other possibilities are allowed as well, if one
has some physical arguments in favor of, say, extending the time
window farther to the next-nearest kink, or a kink of a special type.
\end{itemize}

The update procedures thus defined comprise an ergodic stochastic
process that operates on the entire configuration space of the system.
All configurations, including those with nonzero winding numbers 
and different number of particles, are accounted for.
Extra particles are inserted/removed from the system 
when $a{\dag}_j(\tau_2)$ makes
a complete loop in time (relative to $a_i(\tau_1)$), i.e., 
when $\tau_2-\tau_1$ changes by multiples of $\beta$. Winding numbers are
introduced when $j-i$ changes by multiples of $L$.
The key  point of our approach is that each local update makes a
contribution to the ${\cal G}(i, \tau )$ histogram, except rare
cases in which there are no worms in the configuration; these configurations
contribute to the ``diagonal" (or conventional) statistics of 
closed worldlines. Contrary to the standard calculation, we do not
adjust all degrees of freedom to the current positions of worms, but rather
probe and update the whole configuration through their motion. 
This almost trivial modification results in a factor of $(L^d \beta )$ 
acceleration of the scheme!

It is instructive to draw an analogy between the motion of worldline 
discontinuities and the loop cluster update rules. As is easily 
seen, the basic elements of the single-loop LCU method known as
``optional decay" and ``forced transition" \cite{Beard} 
correspond to a particular
evolution of the worldline discontinuities (``optional decay" 
corresponds to the ``jump" procedure, and ``forced continuation"
to the ``antireconnection" procedure).
A closed loop is obtained after annihilating the pair 
$a$ and $a{\dag}$. Notice, however, that in our scheme 
(i) not only closed loops,
but also all intermediate configurations are physically  meaningful
and are included into the statistics; (ii) nothing is based on the special
structure of the system Hamiltonian; (iii) the update
is always local (it is known 
that acceptance rates for large loops become very small 
when an external magnetic field
in the $Z$-direction is applied to the Heisenberg system
(magnetic field is equivalent to a finite chemical potential in
bosonic language); this problem is simply absent in local schemes).

The statistics of discontinuities in space and time is
given by ${\cal G}(i,\tau )$, i.e., it is defined by the Hamiltonian.
In general, the optimal update scheme depends on the quantity being 
calculated, and thus one might wish to control the statistics
of worldline discontinuities ``at will." This can be easily achieved
by introducing  a fictitious spacetime dependent potential acting
between the ``worm" ends, so that their relative positions  are
now distributed according to the function
\[
{\cal G} (i,\tau )\: {\cal Q} (i, \tau )\;,
\]
where ${\cal Q} (i, \tau )$ is arbitrary. In this way, one can
change the typical size and shape of the loops generated by the
``worm" algorithm.

The scope of the present paper is such that we are unable to discuss
here many important details concerning the practical implementation of our
algorithm (optimal triple-linked storage, particular forms of Eqs.\ 
(\ref{P_acc})-(\ref{P_rem}) for each subprocess, optimal management of 
subprocesses, etc.). Readers interested in these issues are
encouraged to take advantage of our FORTRAN code with comments. The code
is written for the 1D boson Hubbard model \cite{code}.
                \section{Illustrative results}
\label{sec:4}

To demonstrate the advantages of the CTWL algorithm, 
we have calculated properties of the 1D boson Hubbard model
(Eq.~(\ref{Ham}) with $U_{ij}=U_0~ \delta_{ij} $) for various
coupling constants $U_0$ and particle densities $\rho $. 

Comparison with the exact diagonalization results
for small systems has demonstrated the lack of any detectable 
systematic error. In particular, for a system with eight lattice sites
and six bosons, and on-site repulsion $U=0.5$, the exact diagonalization 
result for the ground-state energy is $E_G=-10.49209$, while long-run 
Monte Carlo simulations yield $E_G=-10.4922(2)$, i.e., a result with
relative accuracy better than $10^{-4}$.

It is well known that a commensurate system
with $\rho =1$ undergoes a superfluid -- Mott-insulator
transition of the Berezinskii - Kosterlitz - Thouless 
type \cite{Berezinskii,Kosterlitz} 
when the on-site interaction is strong enough (for the most accurate 
estimate of the transition point $U_0=1.645 t$, see Ref.~\cite{Svistunov}).
In the superfluid phase, including the critical point, one can
utilize knowledge of the long-wavelength behavior of the system.
As explained by Haldane \cite{Haldane}, 
the energy associated with extra particles
and nonzero winding numbers is quadratic in $M$ and $N-\overline{N}$
(for simplicity, in what follows we count particle numbers from
the commensurate value: $N \to N-L$ and  $ \overline{N} \to 
\overline{N}-L$). This means that
the corresponding probability distribution in $M$ and $N$ is
a Gaussian, i.e.,
\begin{eqnarray}
W(N,M) & \propto & \exp \bigg[ -{L \over 2\beta \Lambda_s(0) }M^2
-{\beta \over 2L \kappa(0) }(N-\overline{N})^2 \bigg]  \nonumber 
\\
& \propto & \exp \bigg[ -{\pi K(0) \over 2} \left(
 {L \over c\beta}M^2 -{c\beta \over L} (N-\overline{N})^2  \right) \bigg] \; .
\label{histogram}
\end{eqnarray}
The zero argument of the superfluid stiffness $\Lambda_s$ and 
compressibility $\kappa$ denotes values at $T=0$. Here 
$K^{-1}=\pi \sqrt{\Lambda_s \kappa }$ is the index that controls the
asymptotic behavior of the correlation functions, and $c$ is the 
speed of sound.

At the critical point, $K(L)$, $\kappa (L) $ and $\Lambda_s(L)$ are 
system-size dependent quantities, with $K(L \to \infty ) \to 1/2 $.
Since the speed of sound is unrenormalizable in a homogeneous system, 
it is sufficient to study scaling 
equations for the critical index only. In fact, the solution of the 
renormalization group (RG) equations for $K(L)$ can be ``visualized"
by considering the logarithmic derivative of the Green's function,
since its index is just $K/2$:
\begin{equation}
K(l )= - 2 {d\ln {\cal G}(r) \over d\ln r }\;;\;\;\;\;\;\; l=\ln r\;.
\label{derivative}
\end{equation}
Here we have introduced the variable $r^2 =x^2+(ct)^2 $, which
by conformal invariance describes asymptotic decay of ${\cal G}$ 
both in space and time. 

Expressions (\ref{histogram}) and (\ref{derivative}) allow for a
comprehensive test of the new algorithm.
It is also tempting to consider a large system right at the quantum critical
point and to evaluate its properties under the most unfavorable
conditions for the standard worldline method. 
To calculate the critical index and the speed of sound, we 
considered a ring with $100$ lattice sites and $\beta=100/t$. 
The critical parameters of the Hamiltonian are 
$U_0=1.645 t$ and $\mu = 1.94 t$ \cite{Svistunov}. We had no problems
in accumulating sufficient statistics of winding numbers
and $N$ for this system (the corresponding calculation
is virtually impossible using the standard worldline algorithm).
Simple manipulations with the exponents in (\ref{histogram}) 
result in the following expressions:
\begin{eqnarray}
\overline{N}&=&{N^2-r_N \over 2(N^2+r_N) } \;; \;\;\;\;\;
r_N={\ln [W(0,N)/W(0,0)] \over \ln [W(0,-N)/W(0,0)] }\;; \nonumber \\
\kappa(0)&=&{\beta \over L} \: {N^2 \over p_N} \;; \;\;\;\;\;
p_N=-\ln \bigg[ {W(0,N)W(0,-N) \over W^2(0,0) } \bigg] \;; \nonumber \\
\Lambda_s(0)&=&{L \over \beta } \: {M^2 \over g_M} \;; \;\;\;\;\;
g_M=-\ln \bigg[ {W(0,M)W(0,-M) \over W^2(0,0) } \bigg] \;.
\label{fromgauss}
\end{eqnarray}
If one is interested in evaluating directly $K(0)$ then
\begin{equation}
K(0)={(p_Ng_M)^{1/2} \over \pi \vert NM \vert } \;.
\label{index}
\end{equation}
The choice of $N$ and $M$ here is arbitrary, but
for numeric reasons, the optimal $N$ and $M$ correspond to values
where $(p_N,g_M) \sim 1$. The advantage of working with
nonzero winding numbers in the grand canonical ensemble is obvious:
in a single MC calculation, one collects all the necessary
information about the parameters in the effective long-wavelength 
action, which is very convenient in determining quantum critical points
from $K=K_c$. For the aforementioned system we found 
$c/t=2.4(1)$, and $K(l= \ln (100))=0.47(1)$.

One note is in order here. The Gaussian distribution (\ref{histogram})
implies that the system is in the superfluid phase. In the general case
one has to define the compressibility 
as $\kappa = d \rho /d\mu $, where by definition $\rho = \overline{N}/L$.
The superfluid stiffness $\Lambda_s $ is defined as the coefficient 
relating persistent current and gauge phase when $\varphi \to 0$;
this yields \cite{Pollock} $\Lambda_s = \overline{M^2}L/ \beta$.
 
Finally, we used our method to 
evaluate the Green's function ${\cal G}(i,\tau )$ and to extract 
the critical index of the Berezinskii - Kosterlitz - Thouless 
transition from its asymptotic behavior; 
one can then check the consistency of all calculations.
Since the CTWL simulation yields a two-dimensional 
histogram for ${\cal G}(i,\tau )$, 
much more accurate results for $K(l)$ are obtained by 
computing logarithmic derivatives along different directions in the
$(x,\tau )$ plane with subsequent angular averaging. The speed of sound,
which is necessary for such a calculation, is extracted from the
asymmetry between $x$ and $\tau $ in the asymptotic decay of ${\cal G}$.
For the Green's function calculation, we considered a ring
with $450$ lattice sites and $\beta=200/t$. In Fig.\ 4, we
show the short-range behavior of the Green's function in
the $\tau $-direction, with the characteristic jump at $\tau =0$.
The dashed curve is a linear interpolation between the calculated points.
In Fig. 5, we present the full-scale behavior of ${\cal G}(x,\tau )$
by plotting it as a function of $r=(x^2+(ct)^2)^{1/2}$
along the time and $x=c\tau $ directions. In accordance with conformal 
invariance, for large $r$ the two curves
are indistinguishable to within the statistical errors.
The speed of sound obtained from the Green's function
is $c/t=2.4(1)$, and analysis of the 
logarithmic derivative (\ref{derivative}) yields $K( l= \ln (100) )
=0.46(2)$.

It is worth noting that our calculations
for $W(N,M)$ and ${\cal G}$ were performed on a Pentium-90 PC. 
None of these results (e.g., for $L>100$) can be obtained by other methods,
even with the use of supercomputers.

The strong on-site disorder at low temperatures is a severe trial for 
most Monte Carlo schemes. Cluster methods suffer from inefficient 
global Metropolis updates here, while standard canonical-ensemble algorithms
suffer from slowing down due to one-particle local 
minima in the effective action (the lowest single-particle states are well
localized, and probing different configurations requires deep sub-barrier
motion). The unique feature of our ``worm" update method -- the possibility 
of locally seeding an extra world line at any point
in the spacetime continuum -- obviates these problems. 

To demonstrate the efficiency of our method, we present the results of just
one nontrivial calculation -- the dependence of the average particle number
on the chemical potential in the Bose glass (BG) phase of the 1D disordered 
Hubbard model, Fig.\ 6. We consider a system with $L=60$ sites at 
$\beta=150$, $U=3t$. Disorder is introduced by randomly distributing
the on-site potential $\sum_i \epsilon_i n_i$ 
between $-\Delta$ and $\Delta$, with $\Delta=6t$.
The curve $\langle N \rangle (\mu )$ allows precise 
determination of the low-energy 
quasiparticle spectrum of the system to order $0.01t$,
which is equivalent to the calculation of the total system energy
to a relative accuracy of order $10^{-4}$. The entire plot of
$\langle N \rangle (\mu )$
was obtained in a few days of CPU time on a PentiumPRO-200 processor.

To obtain further evidence of the effectiveness of our method in more 
familiar problems, the reader is referred to a calculation of the
superfluid -- Bose glass -- Mott insulator phase diagram for the 1-D 
disordered boson Hubbard model \cite{PS}.

                \section{Concluding remarks}
\label{sec:5}

Although the CTWL algorithm developed here is quite general, some
aspects deserve special discussion. What if the interaction radius
$r_o$ is large? All of the procedures update configuration fragments
with the typical duration $\tau_r -\tau_l \sim 1/t$. Since
the method is exact, we trace all the kinks within the interaction 
radius, because they contribute to the function
$R(\vec{\tau })$. This means that the interval $(\tau_l,~\tau_r)$
is further split into $N_{\tau_{l,r}} \sim r_o^d z \gg 1$ 
subintervals ($z$ is the coordination number), and each subinterval
requires special consideration. If $r_o$ is as large as the system
size, then the whole scheme is in trouble, becoming a ``victim of
exactness."

The idea of solving such a problem is demonstrated by the stochastic
series expansion method \cite{Sandvik-review,Sandvik91,Sandvik92}.
One might well wonder why continuous-time schemes, 
which contain as an essential ingredient an evaluation of time integrals,
work as efficiently as SSE, which has all these integrals being evaluated
exactly right at the start? Also, why is keeping the
potential energy $U$ in the $\tau$-exponent not at all an advantage
if $U \sim t$? The point is that the MC process is exact only for
asymptotically long computation times, and there is no reason to calculate
anything more precisely than the unavoidable statistical error,
especially if the corresponding calculation becomes the bottleneck
for the whole scheme. Evaluating time integrals in CTWL or 
reproducing exponents by expanding in power series in $U \sim t$
are just two cases that illustrate this point.

Suppose that all particles in the system interact with one another,
so that formally $r_o =L$, but 
$\int d\vec{r}~ \rho (\vec{r}) U(\vec{r} ) = F(L) \ne \infty$.
We divide the interaction Hamiltonian into two parts,
$H_{\mbox{\scriptsize int}}^{(1)}(r<r_t) +
H_{\mbox{\scriptsize int}}^{(2)}(r>r_t)$, by introducing the truncation
radius
\begin{equation}
\int_{\mid r \mid > r_t } d\vec{r} \: \; \overline{\rho } \, U(\vec{r} ) \:
= \: t \;.
\label{radius}
\end{equation}
We then write $H_o=H_{\mbox{\scriptsize int}}^{(1)}$ and combine
$H_{\mbox{\scriptsize int}}^{(2)}$ with $V$ (see Sec.~\ref{sec:2}),
i.e., the long-range part of the interaction Hamiltonian is
now considered to consist of diagonal kinks. Because of the definition 
(\ref{radius}),
the total number of kinks within the time interval $\sim 1/t$ remains
finite and independent of system size. 

The case of a divergent integral $\int d\vec{r} \; \rho (\vec{r}) 
U(\vec{r} )=F(L) \to \infty$
is more subtle, since the number of diagonal kinks within the time
interval $\sim 1/t$, given by $F(L)$, is now large
(if $F(L)$ is a logarithmic function of $L$, we do not
regard this problem as serious). 
On the other hand,
for long-range interactions the so called ``mean-field approximation"
becomes more accurate. Since the mean-field potential is easy to
account for  analytically (and numerically), one now has to deal with
fluctuations, and these quite often satisfy the condition
\begin{equation}
\left| \int d\vec{r}~ (\rho (\vec{r})-\overline{\rho } )~
U(\vec{r} )~ \right| =\delta F(L) \ne \infty \;.
\label{fluct}
\end{equation}
In Appendix A we explain how to organize the Monte Carlo process
using the mean-field approximation for the configuration weight.
The net result is that even for long-range potentials, the
calculation time can remain independent of system size.

In this paper we have concentrated on the Green's function calculation
by restricting the number of worldline discontinuities to $0$ or $2$.
Of course the scheme can be trivially extended to include the case with 
a larger number of discontinuities, if one is interested in the two- or
$n$-particle Green's function or $n$-point vertex. More generally,
our scheme makes it possible to work with Hamiltonians that do not 
conserve the number of particles, i.e., when there are sources with 
finite strength in the bare Hamiltonian.

Although in this paper we consider a system with discrete Hilbert space 
in detail, the principles of update in continuous time developed here are 
much more general. Mathematically, we construct an exact method (in the 
statistical limit) of averaging over a distribution
represented as a series of integrals with an ever-increasing number of
variables, but with essential similarity among the terms of the series,
allowing their local comparison (weighting). We may call such structures 
integrals with a variable number of variables -- 
VNV integrals. Physically, we sum a perturbative expansion in the
interaction picture for some observable of a large but essentially
finite-size system. (For a system with discrete Hilbert space, the only
continuous variables in this expansion are the times of virtual
transitions.) But perturbative expansions for continuous systems
also have the structure of VNV integrals, with additional integrations
over some continuous variables. Thus (apart from the fact that for
spatially continuous systems one cannot expand the kinetic part
of the Hamiltonian and must use the potential energy as a perturbation),
there is no qualitative difference between perturbative expansions for 
continuous and discrete systems. The general method of evaluating
VNV integrals is given by Eqs.\ (\ref{balance}) and 
(\ref{balcond}) -- (\ref{P_rem}), where the vector 
$\vec \tau$ now stands for any set of continuous variables, 
and the function $R(\vec \tau)$ is defined straightforwardly,
given the particular form of the series.

               \section{Acknowledgment}
\label{sec:6}

We would like to thank V. Kashurnikov, A. Sandvik, M. Troyer,
H. Evertz, B. Beard, and N. Kawashima for inspiring discussions 
of existing Monte Carlo schemes and valuable comments on the 
final version of the paper. 
This work was supported by Grant No. INTAS-93-2834-ext (of the
European Community) and partially by the Russian Foundation for Basic
Research (Grant No. 95-02-06191a).

\appendix
\section{Long-range potentials}
Suppose that we are dealing with the case $F(L) \to \infty$, but
finite $\delta F(L)$. The idea is to organize the Monte Carlo process
in such a way, that in most updates we simply ignore fluctuations,
and account for distant particles by replacing them with a homogeneous
density distribution. Obviously, for the scheme to remain accurate,
in some updates we have to consider deviations from the mean-field 
distribution. The goal is to address the procedure dealing with
distant fluctuations with  the  small  probability  which  is  at  least 
inversely
proportional to the number of operations in this procedure.

Consider again the balance equation for the given pair of subprocesses,
but now including the possibility of completing the same
update procedure in a number of ways:
\begin{equation}
A_0 \, p_{\mbox{\scriptsize c}} \, W(\vec \tau) \, 
\sum_{j=0}^{j_*}~\gamma^{(j)} ~ P_{\mbox{\scriptsize acc}}^{(j)} (\vec \tau) 
\, d \vec \tau \; - \; 
dA_n(\vec \tau ) \, p_{\mbox{\scriptsize a}} \, 
\sum_{j=0}^{j_*}~\gamma^{(j)} ~ P_{\mbox{\scriptsize rem}}^{(j)} (\vec \tau) 
 \; = \; 0 \; \; .
\label{balance-A}
\end{equation}
Here $\gamma^{(j)}$ is the probability of using the $j$-th version
of the update procedure. We require $\sum_{j=0}^{j_*}~\gamma^{(j)} =1$
and $\gamma_0\gg \gamma_1 \dots \gg \gamma_{j_*}$. We also
assume that the procedure $j_*$ corresponds to the exact treatment
of all fluctuations. Other quantities have 
exactly the same meaning as in (\ref{balance}).
The self-balance condition now reads (compare Eq.~(\ref{balcond}))
\begin{equation}
W(\vec \tau) \sum_{j=0}^{j_*}~\gamma^{(j)} ~ 
P_{\mbox{\scriptsize acc}}^{(j)} (\vec \tau )  =R(\vec \tau )
\sum_{j=0}^{j_*}~\gamma^{(j)} ~ P_{\mbox{\scriptsize rem}}^{(j)} (\vec \tau ) 
\;.
\label{balcond-A}
\end{equation}

To satisfy (\ref{balcond-A}) we suggest the following scheme.
Let $R^{(j)} (\vec \tau )$ be the distribution corresponding to
the exact treatment of fluctuations up to the distance $r^{(j)}$ with
$r^{(0)} \ll  r^{(1)} \ll \dots \ll r^{(j_*)}=L $, and the mean-field
treatment of more distant ($r>r^{(j)}$) particles. We can write then
\begin{equation}
R^{(j)}=R^{(0)}+\delta R^{(1)} + \dots \delta R^{(j)}\;,
~~~~~R^{(j_*)}\equiv R \;.
\label{distributions}
\end{equation}
If $\delta F$ is finite and $r^{(0)}$
is sufficiently large, then all $\delta R^{(j)}$ are small. We then choose
$\gamma^{(0)} \approx 1 $ and
\begin{equation}
P_{\mbox{\scriptsize acc}}^{(0)} (\vec \tau)  \: = \: \left\{ 
\begin{array}{ll}
R^{(0)} (\vec \tau) / W(\vec \tau) \; ,
\mbox{~~~~~~~~~~ if $R^{(0)} (\vec \tau) < W(\vec \tau) $} \;\; , \\
\;\;\;\;\;\;1 \; ,  \mbox{~~~~~~~~~~~~~~~~~~~~~ otherwise} \;\;\;\;\; ,
\end{array} \right.
\label{PA-0}
\end{equation}
\begin{equation}
P_{\mbox{\scriptsize rem}}^{(0)}(\vec \tau)  \: = \: \left\{ 
\begin{array}{ll}
W(\vec \tau) / R^{(0)} (\vec \tau) \; ,
\mbox{~~~~~~~~~~ if $R^{(0)} (\vec \tau) > W(\vec \tau) $} \;\; , \\
\;\;\;\;\;\;1 \; ,  \mbox{~~~~~~~~~~~~~~~~~~~~~ otherwise} \;\;\;\;\; ,
\end{array} \right.
\label{PR-0}
\end{equation}
and solve the self-balance condition deductively by requiring
\begin{equation}
 W(\vec{\tau} ) \sum_{j=0}^{k} ~ \gamma^{(j)} ~ 
P_{\mbox{\scriptsize acc}}^{(j)} (\vec \tau )  =R^{(k)}(\vec{\tau} )
\sum_{j=0}^{k} ~ \gamma^{(j)} ~ P_{\mbox{\scriptsize rem}}^{(j)} (\vec \tau ) 
\; ,
\label{balcond-A2}
\end{equation}
or equivalently
\begin{equation}
\gamma^{(k)}\big[ W ~P_{\mbox{\scriptsize acc}}^{(k)} - R^{(k)}~
P_{\mbox{\scriptsize rem}}^{(k)} \big] =\delta R^{(k)}
\sum_{j=0}^{k-1}~\gamma^{(j)} ~ P_{\mbox{\scriptsize rem}}^{(j)} \;.
\label{balcond-A3}
\end{equation}
The final answer can be written 
\begin{equation}
P_{\mbox{\scriptsize acc}}^{(k)}  \: = \: \left\{ 
\begin{array}{ll}
\big[ \delta R^{(k)}
\sum_{j=0}^{k-1}~\gamma^{(j)} ~ P_{\mbox{\scriptsize rem}}^{(j)}\big] /
\big[ \gamma^{(k)} W(\vec \tau) \big] \;, 
\mbox{~~~~~ if $\; \delta R^{(k)} > \; 0$} \;\; , \\
\;\;\\
\;\;\;\;\;\;0 \; ,  \mbox{~~~~~~~~~~~~~~~~~~~~~~~~~~~~ otherwise} \;\;\;\;\; ,
\end{array} \right.
\label{acc-A}
\end{equation}
\begin{equation}
P_{\mbox{\scriptsize rem}}^{(k)}  \: = \: \left\{ 
\begin{array}{ll}
-\big[  \delta R^{(k)}
\sum_{j=0}^{k-1}~\gamma^{(j)} ~ P_{\mbox{\scriptsize rem}}^{(j)} \big] /
\big[ \gamma^{(k)} R^{(k)} (\vec \tau) \big] \;, 
\mbox{~~~~~ if $\; \delta R^{(k)} < \; 0$} \;\; , \\
\;\;\\
\;\;\;\;\;\;0 \; ,  \mbox{~~~~~~~~~~~~~~~~~~~~~~~~~~~~~ otherwise} \;\;\;\;\; 
.
\end{array} \right.
\label{rem-A}
\end{equation}
Since all $\delta R^{(k)}$ are assumed to be small, it is 
possible to keep $\gamma^{(k)} \ll 1$ (for $k=1,2,\dots j_*$),
but large enough to avoid situations with 
$P_{\mbox{\scriptsize acc}}^{(k)}>1$ or 
$P_{\mbox{\scriptsize rem}}^{(k)}>1$.


\figure{{\bf FIGURE 1.}
\\
A typical $8$-site configuration with two worldline discontinuities
marked by filled circles.  The width of 
the solid line is proportional to the site occupation number, and
dashed lines are empty sites.
}

\figure{{\bf FIGURE 2.}
\\
Jump procedure for the annihilation operator. a) Initial
configuration fragment; b) suggested variation (in 
the antijump procedure, (b) is the initial configuration and (a) is the
suggested variation).
}

\figure{{\bf FIGURE 3.}
\\
Reconnection procedure for the annihilation operator. a) Initial
configuration fragment; b) suggested variation (in 
the antireconnection procedure, (b) is the initial configuration and 
(a) is the suggested variation).
}

\figure{{\bf FIGURE 4.}
\\
Short-time behavior of the Green's function 
${\cal G}(0,\tau )$ of the commensurate 1-D Hubbard model at the 
quantum critical point.
}

\figure{{\bf FIGURE 5.}
\\
Long-range behavior of ${\cal G}(i,\tau )$, demonstrating
conformal invariance.
}

\figure{{\bf FIGURE 6.}
\\
Number of particles vs.\ chemical potential in a large Bose-glass 
cluster at macroscopically low temperature.
}

\newpage
 
               

\begin{thebibliography}{99}
\bibitem{com1} In particular, while LCU is formulated explicitly for
arbitrary spin $S$ \cite{KG,Kawashima96}, each spin number requires a
separate code, with the complexity of the code growing drastically with
$S$ (e.g., for $S=1$ the basis set already contains 105 graphs
\cite{Kawashima96} !). To the best of our knowledge, 
the LCU codes written up to now deal only with $S<2$.
\bibitem{com2} To stress this important point, we find it reasonable to
distinguish between ``efficient" LCU algorithms and others.
By ``efficient" LCU we mean algorithms in which 
detailed balance is taken care of in the cluster-building rules, 
not in having a global Metropolis step with small acceptance
rates for large clusters. Nevertheless, in certain cases ``inefficient"
LCU code works reasonably well \cite{Kawashima94}.
\end{thebibliography}
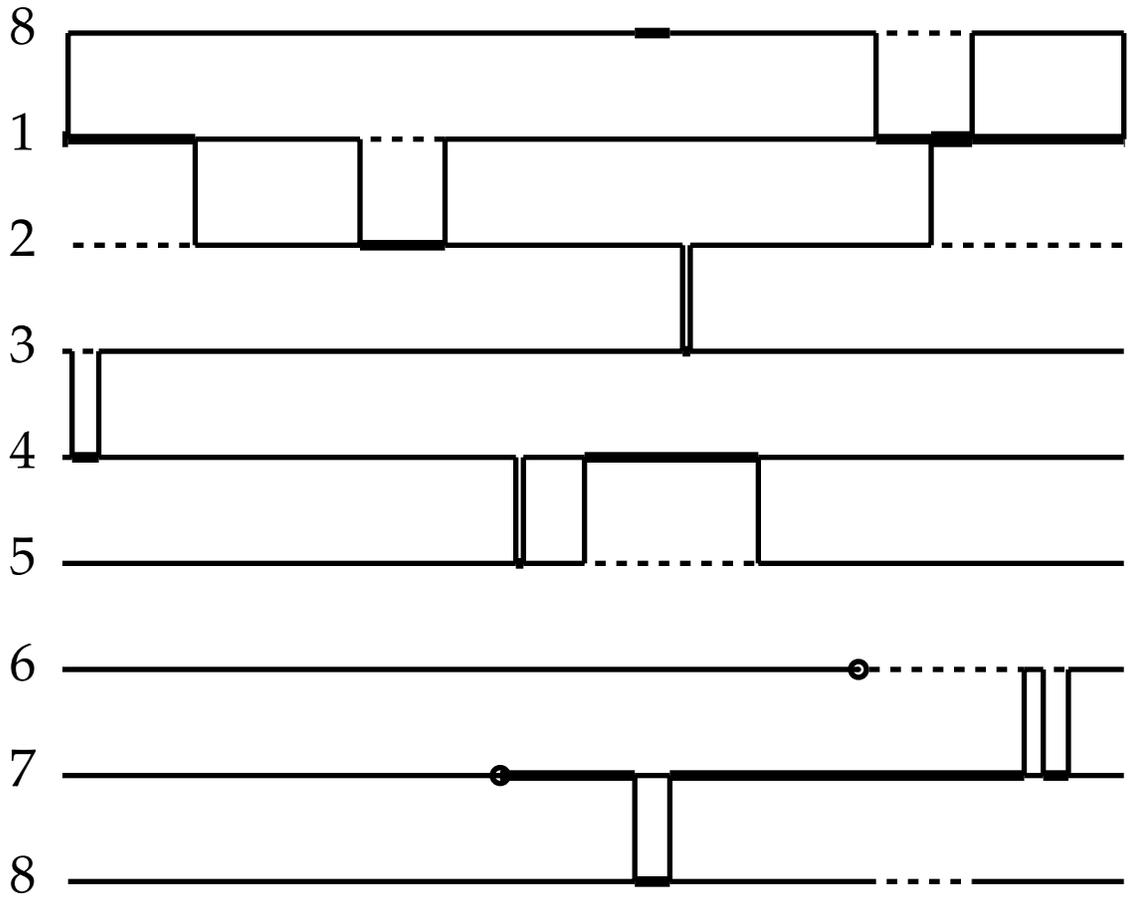
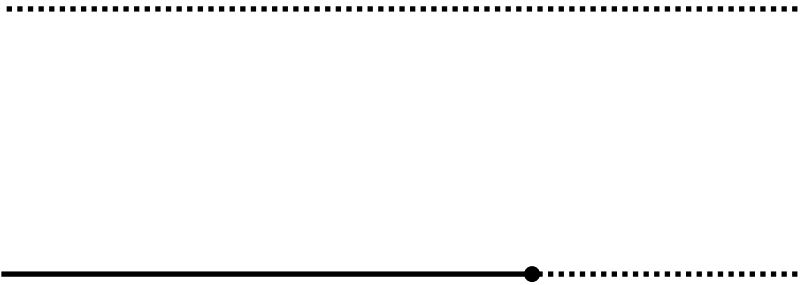
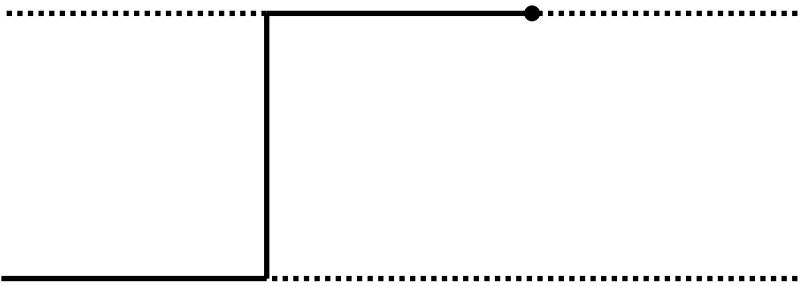
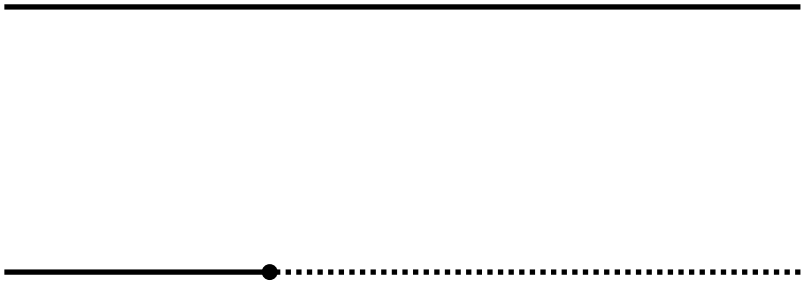
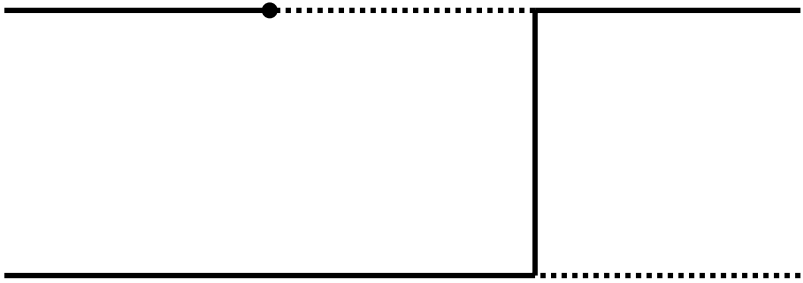
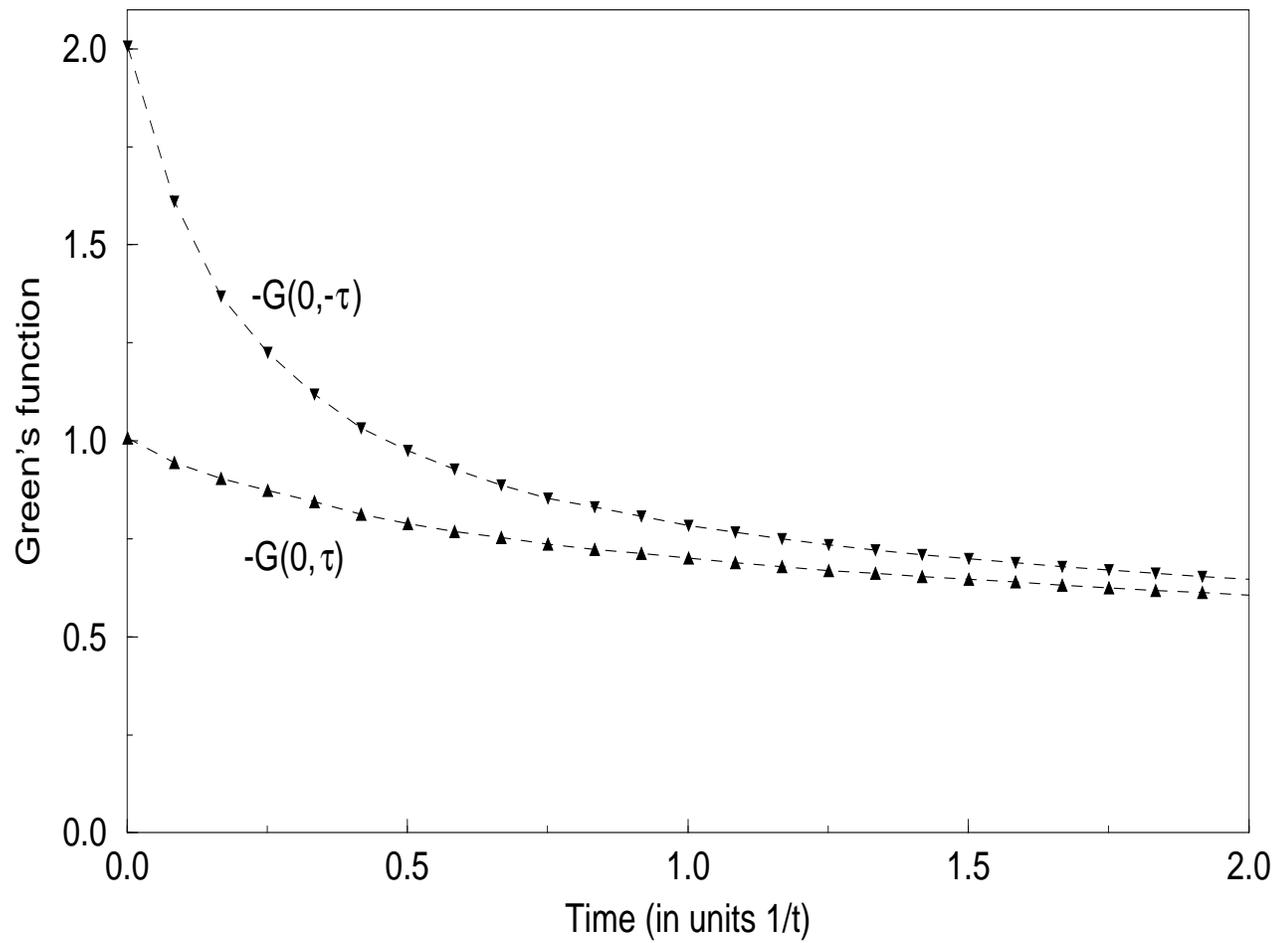
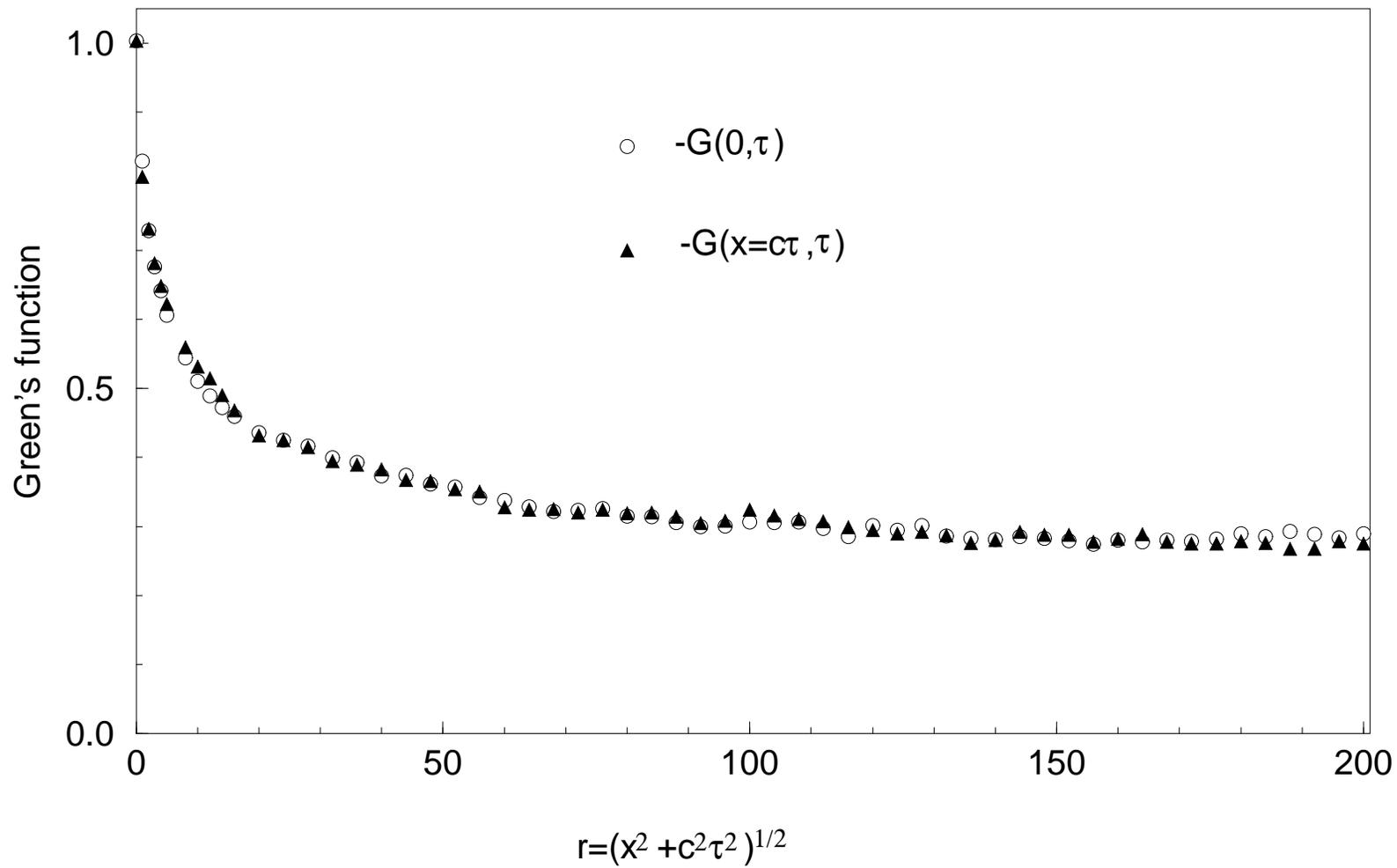
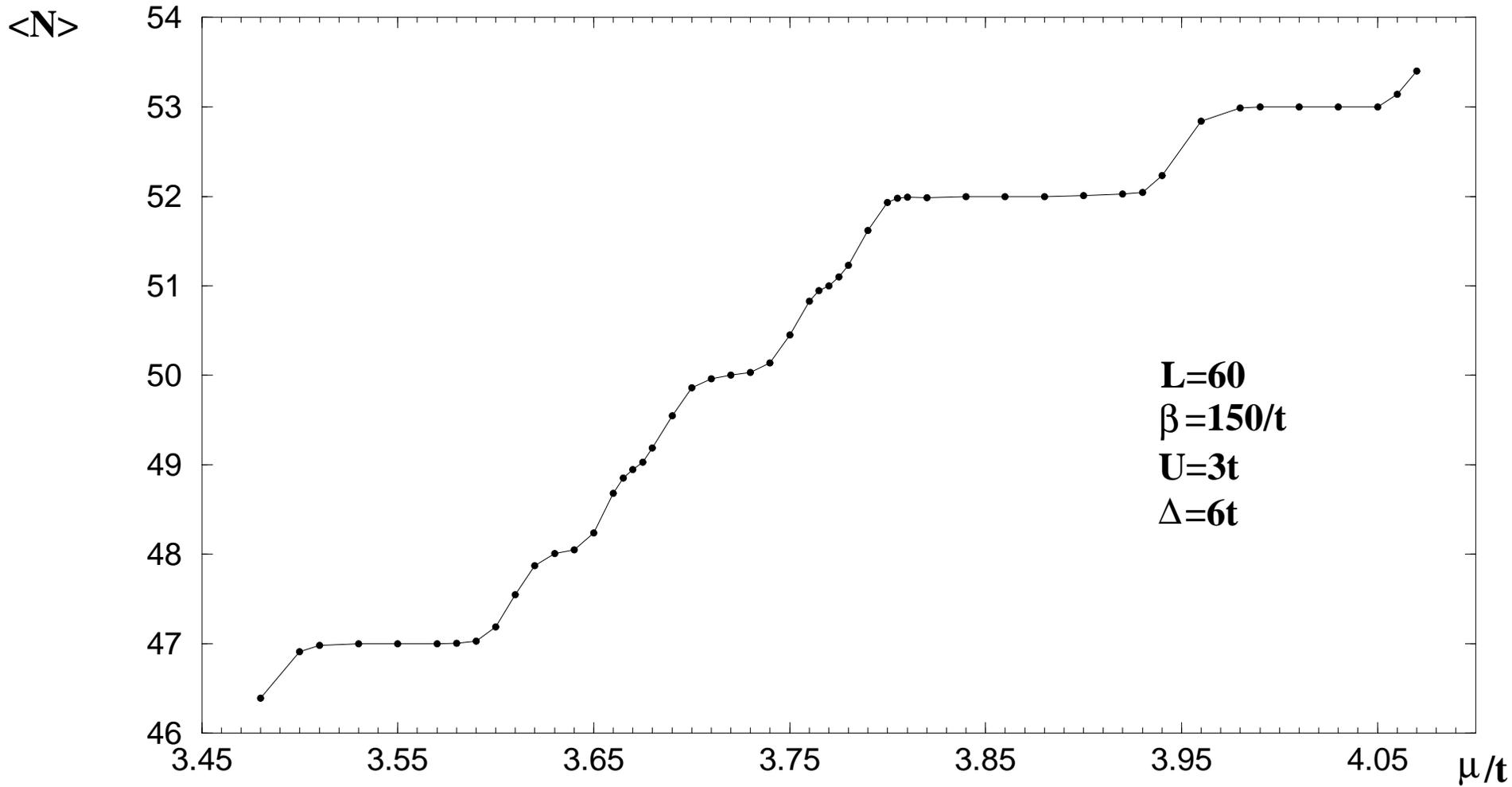
\end{document}